\documentstyle[pra,aps]{revtex}
\begin{document}
\draft
\author{V.B. Svetovoy and M.V. Lokhanin}
\address{{\small Department of Physics, Yaroslavl State University,} \\ {\small %
Sovetskaya 14, Yaroslavl 150000, Russia}}
\title{Precise calculation of the Casimir force between gold surfaces}
\date{\today }
\maketitle

\begin{abstract}
We analyse the result of precise measurement of the Casimir force between
bodies covered with gold. The values of the parameters used to extrapolate
the gold dielectric function to low frequencies are very important and
discussed in detail.  The finite temperature effect is shown to exceed
considerably the experimental errors. The upper limit on the force is found
which is smaller than the measured force. Many experimental and theoretical
uncertainties were excluded with gold covering and we conclude that,
possibly, a new force has been detected at small separations between bodies.
\end{abstract}

The Casimir force \cite{Casimir} attracts considerable attention nowadays.
At early stage it has been measured between two conducting plates \cite{Spar}
with large errors and much more accurately for plate and sphere covered with
chromium \cite{BO}. A new wave of interest raised when the force between
metallized bodies has been measured with high precision \cite{Lam1,MR,RLM}.
Theoretical prediction for the force is not so straightforward as expected.
In the original Casimir work \cite{Casimir} the force $F^{pl}(a)=\frac{\pi ^2%
}{240}\frac{\hbar c}{a^4}$ has been found as change in the zero point energy
between two parallel perfectly conducting plates separated by the distance $%
a $. For sphere and plate the expression above has to be modified with the
proximity force theorem (PFT) \cite{PFT} to the following

\begin{equation}
\label{F0}F_c^0(a)=\frac{\pi ^3\hbar c}{360}\frac R{a^3}, 
\end{equation}

\noindent where $R$ is the sphere radius. Real metals are not ideal
conductors and there is an important correction to (\ref{F0}) taking into
account their real properties. In the Lifshitz theory \cite{Lif}, which
generalizes the Casimir approach, the force is represented in terms of the
metal dielectric function $\varepsilon \left( i\omega \right) $ at imaginary
frequencies. This theory was applied for description of the experiments in 
\cite{RLM,Lam2}. Different but equivalent method has been developed in \cite
{LR1}, where detailed calculations were made for $Al$, $Au$, and $Cu$
metallization of the bodies using the handbook optical data for these
metals. The result for $Au$ is in good agreement with the torsion pendulum
experiment \cite{Lam1}.

For the atomic force microscope (AFM) experiment \cite{MR,RLM} $Al$
metallization has been used. This metal is oxidized easily and charges can
be trapped in the oxide. To prevent influence of these charges on the force, 
$Al$ was covered with a thin layer ($20$ or $8\ nm$ thick) of $Au/Pd$. The
conclusion has been made \cite{RLM} that the result of the AFM experiment
was in agreement with the theory. However, in all calculations \cite
{MR,RLM,KRMM} the role of $Au/Pd$ layer was ignored. Importance of this
layer was stressed in \cite{SL1} (see also \cite{SL3}), where the upper
limit on the force has been found using parameters of single-crystalline
materials. This limit was smaller than the measured force in both
experiments \cite{MR,RLM} and the discrepancy far exceeded the experimental
errors. To exclude uncertainties connected with aluminum oxidation, $Au/Pd$
layer and clarify the origin of the discrepancy, it was proposed \cite{SL1}
to use $Au$ metallization instead of $Al$. Very recently the AFM experiment
with $Au$ metallization has been carried out \cite{HCM}. In this experiment
the residual potential was reduced to a negligible level so as the surface
roughness. The force was measured for smaller separations up to $62\ nm$
with uncertainty $3.5\ pN$. In the previous experiments the photodiods
response on the cantilever deflection was calibrated assuming that the
plate-sphere system on the contact behaves as a rigid body. As the new
experiment showed this natural assumption was not quite correct and
different calibration method was proposed \cite{HCM}. To all appearance the
excessive force revealed in the experimental data \cite{SL1} originates
mainly from the calibration effect.

The new experiment gives us a solid ground for verification of the theory.
Though good agreement with the theory is reported \cite{HCM}, this
conclusion cannot be considered as final because of the following reasons.
Deviations between theory and experiment were significant at small
separations but the authors averaged them over the whole range of
separations. An error was admitted in extrapolation of the optical data for $%
Au$ to low frequencies. At last, the finite temperature correction was
neglected but it is important at small separations \cite{SL1,SL2}. These are
the main points we will discuss in this paper.

We start from the general expression for the Casimir force given by Lifshitz 
\cite{Lif} modified for the case of sphere-disk geometry:

\begin{equation}
\label{shpl}F(a)=-\frac{kTR}{c^2}{\sum\limits_{n=0}^\infty {}}^{\prime
}\zeta _n^2\int\limits_1^\infty dpp\ln \left[ (1-G_1)(1-G_2)\right] , 
\end{equation}

\noindent where

$$
G_1=\left( \frac{p-s}{p+s}e^{-p\zeta _na/c}\right) ^2,\ G_2=\left( \frac{%
\varepsilon p-s}{\varepsilon p+s}e^{-p\zeta _na/c}\right) ^2,\quad 
$$

\begin{equation}
\label{defin1}s=\sqrt{\varepsilon \left( i\zeta _n\right) -1+p^2},\quad
\zeta _n=2\pi nkT/\hbar . 
\end{equation}

\noindent $\varepsilon \left( i\zeta _n\right) $ is the dielectric function
of $Au$ at imaginary frequencies. The prime over the sum sign indicates that
the first $n=0$ term has to be taken with the factor $1/2$. Special care
needs to treat this term. The formal reason is that $\zeta _n^2$ becomes
zero but the integral over $p$ diverges. The physical reason is that this
term corresponds to the static limit when for metallic bodies we should take 
$\varepsilon \rightarrow \infty $. The resulting contribution in the force
corresponds to the classical limit $F_{cl}\left( a\right) $ for metals \cite
{Temp}

\begin{equation}
\label{clas}F_{cl}\left( a\right) =\frac{kTR}{4a^2}\zeta \left( 3\right) , 
\end{equation}

\noindent where $\zeta \left( N\right) $ is the zeta-function. If the metal
is ideal, this term is canceled by the rest ones in the sum leaving only
negligible temperature correction. The cancelation is incomplete for real
metals \cite{SL2} and the correction becomes important. To evaluate the
Casimir force, the integral instead of the sum in (\ref{shpl}) was used in 
\cite{HCM}. The difference between the sum and integral is just the
temperature correction $\Delta _TF$ which is nearly linear in $T$. The
relative value of $\Delta _TF$ becomes smaller for smaller separations but
its absolute value increases. To have an estimate for the correction, we
used the Drude approximation for $\varepsilon \left( i\zeta \right) $ with
the parameters of perfect single-crystalline gold \cite{SL1} $\omega
_p=1.37\cdot 10^{16}\ s^{-1}$, $\omega _\tau =3.7\cdot 10^{13}\ s^{-1}$ and
found for $\Delta _TF$ the values $4$ and $14\ pN$ at $a=100$ and $60\ nm$,
respectively. Obviously this correction cannot be ignored in precise
calculations.

The function $\varepsilon \left( i\zeta _n\right) $ cannot be measured
directly for a given material but can be expressed via imaginary part of the
dielectric function on the real axis $\varepsilon ^{\prime \prime }\left(
\omega \right) $ with the dispersion relation

\begin{equation}
\label{disp}\varepsilon \left( i\zeta \right) -1=\frac 2\pi
\int\limits_0^\infty d\omega \frac{\omega \varepsilon ^{\prime \prime
}\left( \omega \right) }{\omega ^2+\zeta ^2}. 
\end{equation}

\noindent Actual uncertainty in the force calculations arise from
uncertainties in the function $\varepsilon ^{\prime \prime }\left( \omega
\right) $. Part of the available data on $\varepsilon ^{\prime \prime
}\left( \omega \right) $ is shown in Fig.\ \ref{1}. The solid line
represents the data taken from \cite{Handb1} which were extrapolated to low
frequencies as discussed in \cite{LR1} (see below). The points marked with
the open squares and triangles are taken from \cite{Zol}. The circles
represent the data from \cite{JC}. The points marked with the crosses \cite
{Zol} are the only data in the figure for the evaporated gold film while all
the rest are for the bulk material. The data can be separated conventionally
in two frequency regions with the boundary at the curve minimum $\omega
_1=3.2\cdot 10^{15}\ s^{-1}$. In the region $\omega <\omega _1$ the electron
scattering on defects dominates in the material absorptivity. That is why in
this range one can find different values for $\varepsilon ^{\prime \prime
}\left( \omega \right) $ corresponding to the samples prepared in different
conditions. At $\omega \ll \omega _1$ the dielectric function is well
described by the Drude model

\begin{equation}
\label{Drudr}\varepsilon \left( \omega \right) =\varepsilon ^{\prime
}+i\varepsilon ^{\prime \prime }=1-\frac{\omega _p^2}{\omega \left( \omega
+i\omega _\tau \right) }. 
\end{equation}

\noindent The parameters one can find by fitting the data for the real and
imaginary parts of $\varepsilon \left( \omega \right) $ with (\ref{Drudr}).
It has been done \cite{SL1,SL3} for the points marked by squares and
triangles (rows 2 and 3 in Table \ref{tab1}). Different procedure has been
used for the solid line \cite{LR1}. The data in \cite{Handb1} are available
for frequencies $\hbar \omega >0.1\ eV$. They have been extrapolated to
lower frequencies according to (\ref{Drudr}). The plasma frequency was
estimated using its relation with the concentration $n$ of free electrons
and supposing that every atom produce one electron: $\omega _p=\sqrt{%
e^2n/m_e^{*}\varepsilon _0}$. Here $e$ is the electron charge, $\varepsilon
_0$ is the free space permittivity, and $m_e^{*}$ is the effective electron
mass which is close but larger than $m_e$. $\ $Then at fixed $\omega _p$ the
data \cite{Handb1} have been fitted with (\ref{Drudr}) to find $\omega _\tau 
$ (row 1 in Table \ref{tab1}). The $\omega _p$ found in this way has
actually the largest of possible values and, of course, the real material
has smaller $\omega _p$ due to presence of the defects.

In contrast to low frequencies all the points at $\omega >\omega _1$
describe practically the same curve. That is because at high frequencies the
interband absorption dominates and the defects become unimportant.
Therefore, one can consider this part of the curve as the universal one
(sample independent). We estimated uncertainty in $\varepsilon \left( i\zeta
\right) $ due to a reasonable variation of this part of the curve. It is
less than 1\% for important frequencies $\zeta \sim c/2a$. The resulting
uncertainty in the force will be much less than the experimental errors. One
can separate contributions in $\varepsilon \left( i\zeta \right) $ from
different frequency regions and write $\varepsilon \left( i\zeta \right)
=1+\varepsilon _1+\varepsilon _2+\varepsilon _3$. The contribution of low
frequencies $0<\omega <\omega _0$, where the optical data are unavailable,
can be expressed analytically using (\ref{Drudr}) in the dispersion relation:

$$
\varepsilon _1\left( i\zeta \right) =\frac 2\pi \frac{\omega _p^2}{\zeta
^2-\omega _\tau ^2}\left[ \tan ^{-1}\left( \frac{\omega _0}{\omega _\tau }%
\right) -\frac{\omega _\tau }\zeta \tan ^{-1}\left( \frac{\omega _0}\zeta
\right) \right] . 
$$

\noindent $\varepsilon _2$ and $\varepsilon _3$ can be found numerically
using the data in the ranges $\omega _0<\omega <\omega _1$ and $\omega
>\omega _1,$ respectively. To get an idea of the contributions of different
intervals, we have found for the solid line in Fig. \ref{1} $\varepsilon
_1=26.6$, $\varepsilon _2=8.3$, and $\varepsilon _3=5.3$ at $\zeta =c/2a$
with the smallest separation $a=63\ nm$ (the first point presented in \cite
{HCM}). Let us stress once more that $\varepsilon _3$ does not depend on the
defect concentration in the material while $\varepsilon _1$ is quite
sensitive to this parameter via actual values of $\omega _p$ and $\omega
_\tau $ and its contribution dominates in $\varepsilon \left( i\zeta \right) 
$.

Proceeding exactly as in \cite{LR1} we successfully reproduced $\varepsilon
\left( i\zeta \right) $ given there in Fig.\ 2. However, we failed to
reproduce the same function in \cite{KMM} though the authors claimed to use
the same data and the same values for $\omega _p$ and $\omega _\tau $. The
difference between the curves presented in \cite{KMM} and \cite{LR1} is 46\%
at $\zeta =10^{12}\ s^{-1}$ and decreases with frequency increase. To
understand the origin of this difference, consider the limit $\zeta
\rightarrow 0$ when the dielectric function behaves as $\varepsilon \left(
i\zeta \right) $ $\rightarrow \left( \varepsilon _0\rho \zeta \right) ^{-1}$%
. Here $\rho $ is the static resistivity defined via the Drude parameters as 
$\rho =\omega _\tau /\varepsilon _0\omega _p^2$. The parameters $\omega _p$
and $\omega _\tau $ used in \cite{LR1} correspond to $\rho =3.2\ \mu \Omega
\cdot cm$. For unknown reason the function $\varepsilon \left( i\zeta
\right) $ presented in \cite{KMM} at low frequencies agrees much better with 
$\rho =2.3\ \mu \Omega \cdot cm$ that has no relation with the data in \cite
{Handb1}.

The Casimir force was evaluated in \cite{LR1} and \cite{KMM} without the
temperature correction using the integral instead of the sum in (\ref{shpl}%
). To check the procedure, we repeated the calculations there at $a=100\ nm$
and found for the reduction factor $\eta =F/F_{c}^0$ the values $0.547$ and $%
0.559$ to be compared with $0.55$ \cite{LR1} and $0.56$ \cite{KMM},
respectively. The small difference in $\eta $ results from different
resistivities used to extrapolate the optical data. At $a=63\ nm$ this
difference is not so harmless: we found for the force $F=467\ pN$ and $477\
pN$, respectively. The latter value is in agreement with the theoretical
curve in \cite{HCM} since the calculations have been done as in \cite{KMM}.
The force has to be compared with the experimental value of $491\ pN$. The
difference is 4 times larger than the experimental errors although the rms
deviation of the experiment from theory for the whole range of separations
is only $3.8\ pN$ \cite{HCM}.

Now let us take into consideration the temperature correction which is
positive and will improve agreement with the experiment. We make
calculations using Eq.(\ref{shpl}) with $\varepsilon ^{\prime \prime }\left(
\omega \right) $ given by the solid line, squares, and triangles in Fig. \ref
{1}. For the squares there is a gap in the interval $6.3\cdot 10^{14}\
s^{-1}<\omega <\omega _1$ which is filled by the linear interpolation on a
log-log scale. Different ways to fill the gap lead to uncertainty in the
force $\pm 1\ pN$. The results of calculations are presented in Table \ref
{tab1}. Note that the force given in \cite{HCM} accidentally coincides with
the right one (row 1) because the error in $\rho $ was compensated by the
temperature correction. The parameters in row 2 are quite close to those of
perfect single-crystal. Therefore, the force has to be maximal in this case 
\cite{SL1} due to maximal reflectivity. However, we see that it is a bit
larger for row 1. The reason is that $\omega _p$ in \cite{LR1} was not
extracted from the optical data but set to be maximal by hand. Real material
will have smaller $\omega _p$ that will lead to smaller force. The residual
force defined as $\Delta F\left( a_i\right) =F_{exp}\left( a_i\right)
-F_{theor}(a_i)$, where $a_i$ are the experimental separations, is shown in
Fig. \ref{2} for $a_i<103\ nm$. It has been calculated for the squares in
Fig. \ref{1} and can be considered as the minimal residual force. As we can
see $\Delta F$ exceeds 4 standard deviations in a few points of the closest
approach. Parameters of evaporated gold films will deviate significantly
from those of the perfect single-crystal because the films contain much more
defects than the bulk material \cite{SL3}. Therefore, the residual force for
real body covering will be definitely larger and will be significant at
larger separations. One can easily improve the upper limit on the predicted
Casimir force by measuring the resistivity of the gold films used in the
experiment.

The experiment \cite{HCM} has been substantially improved in comparison with 
\cite{MR,RLM} and many uncertainties have been excluded. If there is no an
additional systematic error at small separation, the deviation between
theory and experiment has to be considered seriously. In this connection it
is very important to reproduce the result in independent measurements. Good
potential has the micromechanical resonator experiment \cite{CFOR} which
already has the necessary sensitivity but an additional work has to be done
to control the separations $\sim 100\ nm$.

One can speculate that the residual force is explained by a new Yukawa force
mediated by a light scalar boson. In this case interaction of two atoms is
described by the potential

\begin{equation}
\label{VY}V_Y\left( r\right) =-\alpha N_1N_2\frac{\hbar c}r\exp (-r/\lambda
), 
\end{equation}

\noindent where $\alpha $ is the dimensionless interaction constant, $%
\lambda $ is the Compton wavelength of a particle responsible for the
interaction, and $N_{1,2}$ is the number of nucleons in atoms of the
interacting bodies. With 95\% cl the residual force is larger than $%
17-2\cdot 3.5=10\ pN$. Integrating (\ref{VY}) for the bodies in the
experimental configuration one finds the lower limit on the interaction
constant

\begin{equation}
\label{restr}\alpha >\frac{6.27\cdot 10^{-25}e^{a/\lambda }}{%
1-1.74e^{-h/\lambda }+0.75e^{-2h/\lambda }}\left( \frac{100\ nm}\lambda
\right) ^3, 
\end{equation}

\noindent where $a=63\ nm$ is the minimal separation and $h=96\ nm$ is the
gold film thickness. The interaction constant is restricted from the He
burning stars \cite{Raf} as $\alpha <1.5\cdot 10^{-22}$. Then the allowed
region for the wavelength is $\lambda >33\ nm$ or for the scalar boson mass $%
m<38\ eV$.

In \cite{Hagen} it is stated that the sphere-plate geometry cannot be used
for verification of the Casimir effect since a rigorous theoretical
calculation has never been carried out for this configuration. In this
respect we would like to note that such a logic does not allow to verify the
effect for finite-size plates, either, because the plates always can be
considered as parts of spherical surfaces. From physical point of view it is
obvious that the force for infinite plates will be a good approximation if $%
L\gg a$, where $L$ is the plate size. The same is true for a sphere and
plate if $R\gg a$. One can discuss the applicability of PFT but there is no
any reason to worry (see discussion in \cite{KMM} and \cite{SL3}).

When this work was finished we become aware about the paper \cite{BS}, where
the authors claimed that the Schwinger prescription \cite{Temp} for the $n=0$
term in (\ref{shpl}) cannot be applied for nonideal metals. As the result
they found the force in the classical limit which is 2 times smaller than (%
\ref{clas}). We think that the statement needs further investigation (see,
for example, \cite{LamT}), but in any case our conclusion will not change
since the Casimir force will be only smaller for the smaller $n=0$ term.

In conclusion, we found the upper limit on the Casimir force between gold
covered plate and sphere using the handbook data for $\varepsilon \left(
\omega \right) $. The effect of nonzero temperature is shown to be
considerably larger of the experimental errors. The excessive force found at
small separations exceeds 4 standard deviations. The lower limit on the
interaction constant is found assuming that the residual force is the
manifestation of a new Yukawa interaction.

We thank R. Onofrio for attracting our attention to the papers \cite{Hagen}, 
\cite{BS} and for stimulating discussion. We thank also N.V. Mikheev for
discussion and comments.

\begin{figure}

\caption{Imaginary part of the dielectric function 
$\varepsilon^{\prime\prime}(\omega)$ for gold.
The data were collected from different sources (see explanations
in the text).} 

\label{1} 

\end{figure}

\begin{figure} 

\caption{The residual force as a function of separation calculated 
for the squares in Fig. \protect \ref{1}.}

\label{2} 

\end{figure}

\begin{table}   
     \begin{tabular}{|c|c|c|c|c|} 
              & $\omega_p \cdot 10^{-16} $  &  $\omega_{\tau} %
     \cdot 10^{-13} $  &  $\rho\ [\mu \Omega \cdot cm]$  &%
     $F(a_{min})\ [pN]$ \\ \hline
     1       & 1.38 & 5.38  & 3.19  & 477 \\
     2       & 1.37 & 4.06  & 2.44  & 474 \\
     3       & 1.28 & 3.29  & 2.27  & 459 \\      \end{tabular}
  \caption{The Drude parameters, resistivity, and the Casimir force at
  $a_{min}=63 \ nm$ for $\varepsilon^{\prime\prime}(\omega)$  represented
  by the solid line (1), squares (2), and triangles (3) in 
  Fig. \protect\ref{1}.}  %

\label{tab1}
\end{table}

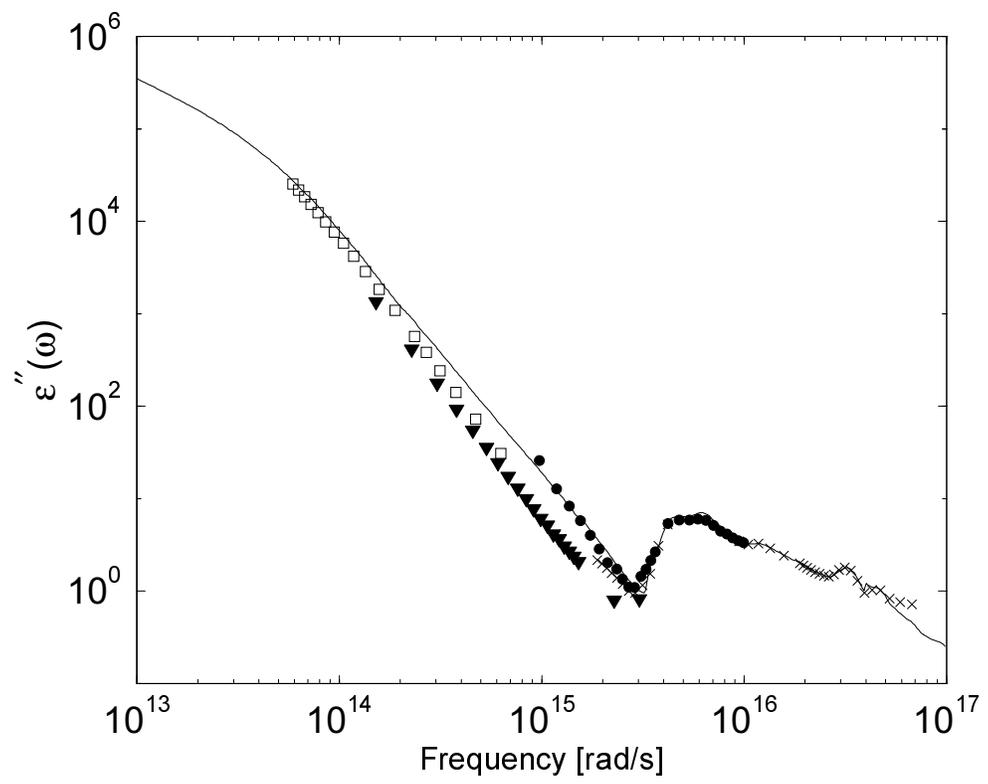
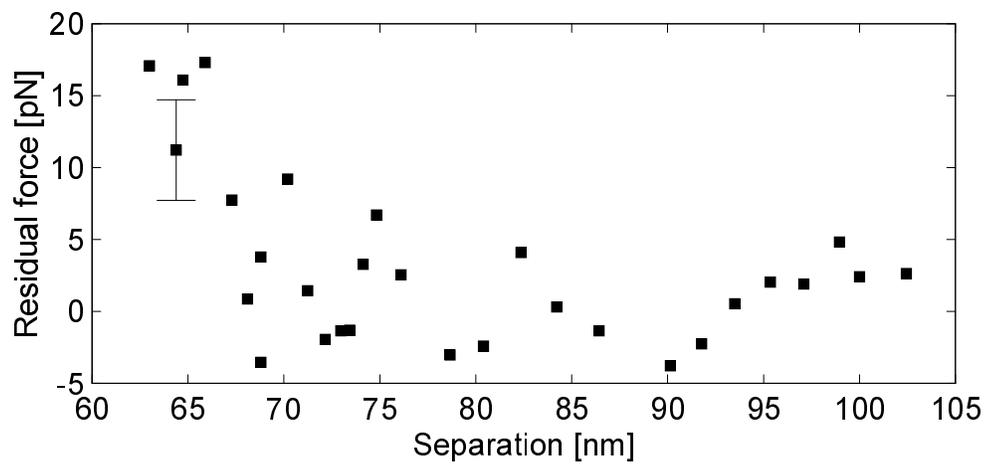
\end{document}